\documentclass[showkeys,showpacs,amsmath,superscriptaddress,amsfonts,preprintnumbers,amssymb,bm,10pt,twocolumn]{revtex4}
\usepackage{graphicx}
\usepackage{bm}
\usepackage{color}
\usepackage{subfigure}
\usepackage{amssymb}
\begin{document}

\title{Shannon entropies and Fisher information of K-shell  electrons of neutral atoms}
\author{Golam Ali Sekh}
\email{skgolamali@gmail.com}
\affiliation{Department of Physics, B.B.College, Asansol 713303, India}
\author{Aparna Saha}
\affiliation{Department of Physics,  Visva-Bharati University, Santiniketan 731235, India}
\author{Benoy Talukdar}
\affiliation{Department of Physics, Visva-Bharati University, Santiniketan 731235, India}
\begin{abstract}
We represent the two K-shell electrons of neutral atoms by Hylleraas-type wave function which fulfils the exact behavior at the electron-electron and electron-nucleus coalescence points and, derive a simple method to construct expressions for single-particle position- and momentum-space charge densities, $\rho(\boldsymbol{r})$ and $\gamma(\boldsymbol{p})$ respectively. We make use of the results for  $\rho(\boldsymbol{r})$ and $\gamma(\boldsymbol{p})$  to critically examine the effect of correlation on bare (uncorrelated) values of Shannon information entropies ($S$) and of Fisher information ($F)$ for the K-shell electrons of atoms from helium to neon. Due to inter-electronic repulsion the values of the uncorrelated Shannon position-space entropies are augmented while those of the momentum-space entropies are reduced. The corresponding Fisher information are found to exhibit opposite behavior in respect of this. Attempts are made to provide some plausible explanation for the observed response of $S$ and $F$ to electronic correlation.
\end{abstract}
\keywords{K-shell electrons; Correlated wave function; Shannon entropy; Fisher information; Correlation effect}
\pacs{03.65.-w; 03.65.Ta; 03.65.Xp; 05.30. - d}
\maketitle
\section{Introduction}
Both information entropy of Shannon \cite{ref1}  and Fisher information \cite{ref2} are characterized by probability density  corresponding to changes in some observable.  Shannon entropy ($S$) is very insensitive to changes in the distribution over a small-sized region and thus possesses a global character. On the other hand, Fisher information ($F$) can detect local changes in the distribution. Consequently, these two information measures give complementary descriptions of disorder in the system. From mathematical point of view the former is a convex and latter is a concave \cite{ref3}. When one grows, the other diminishes. In applicative context it will, therefore, be quite interesting to examine how $S$ and $F$ respond to important physical effects, like the electron-electron correlation, which plays a role in the physics of many-electron systems such as atoms, molecules and solids. There exists a large number of studies \cite{ref4} which attempted to establish that correlation plays a more dominant role in modifying the bare values of $S$ than it does in correcting the Hartree-Fock binding energies of atomic systems. In particular, by working with a few ions in the Helium iso-electronic sequence, Romera and Dehesa \cite{ref5}  introduced the product of the position-space Shannon entropy power and Fisher information as correlation measure of two-electron systems. However, it appears that there is hardly any work to investigate the effect of correlation on the bare values of $F$. Also we note that almost all results for information entropies of correlated systems have been presented for two-electron ionic systems.  The present paper is an attempt to provide a detailed analysis for the effect of electron-electron correlation on both $S$ and $F$ of the K-shell electrons of neutral atoms from Helium to Neon by using an entangled wave function parameterized by Mitnik and Miraglia \cite{ref6}.  The binding energies of K-shell electrons are likely to be significantly affected by the inter-electronic repulsion which gives rise to the so-called Coulomb correlation. It will, therefore, be important to examine the role of correlation in modifying the bare values of $S$ and $F$. This study might assume further significance since except for helium, the K-shell electrons of other atoms will be affected by the presence of passive outer electrons.

In Section $2$ we present the basic equations for computing results for Shannon entropy and Fisher information for many-electron systems and note that both of these are characterized by the associated single-particle charge densities. Since we are interested  in $S$ and $F$ of the active K-shell electrons, we begin Section $3$ by deriving  an analytic approach to construct a single-particle wave function from the separable representation of the two-particle system such that the former gives the diagonal elements of the density matrix \cite{ref7}  appearing in the results for information theoretic quantities. We subsequently adapt the approach to deal with the explicitly $r_{12}=|\boldsymbol{r}_1-\boldsymbol{r}_2|$  dependent non-separable two-particle correlated wave function as given in ref. {6}. The single-particle wave function obtained in this way can be used to evaluate $S$ and $F$ in the position space. But we need the Fourier transform of the constructed position-space wave function to study properties of $S$ and $F$ in the momentum-space. The required transform can be taken by making judicious use of the standard integrals \cite{ref8}
\begin{subequations}
\begin{eqnarray}
\int \frac{1}{\xi}\,\,e^{-\alpha\, \xi+i\, \boldsymbol{\beta}\cdot \boldsymbol{\xi}}\,\,d \boldsymbol{\xi}=\frac{4\pi}{\alpha^2+\beta^2}
\label{eq1a}
\end{eqnarray} 
and
\begin{eqnarray}
\int e^{-\alpha\, \xi+i\, \boldsymbol{\beta}\cdot \boldsymbol{\xi}}\,\,d \boldsymbol{\xi}=\frac{8\,\pi\alpha}{\left(\alpha^2+\beta^2\right)^2}.
\label{eq1b}
\end{eqnarray}
\end{subequations}
We devote Section $4$ to present results for the position- and momentum-space Shannon entropy and Fisher information for the K-shell electrons of atoms from Helium to Neon and thus try to gain some physical weight for the effect of correlation on the bare values of $S$ and $F$. The observed variation of $F$ with atomic number $Z$ and its modification by the action of electron-electron repulsion tend to present some points of contrast with similar changes in the values of $S$. We infer that the opposite behavior for response of $S$ and $F$ to electronic correlation is of purely statistical origin. 

\section{Shannon entropy and Fisher information for many-electron systems}
For a many-electron atom, the position-space Shannon entropy is written as \cite{ref1}
\begin{eqnarray}
S_{\rho}=-\int \rho(\boldsymbol{r})\, \ln \rho(\boldsymbol{r})\,d\boldsymbol{r},
\label{eq2}
\end{eqnarray} 
where
\begin{eqnarray}
\rho(\boldsymbol{r})=\int \left| \psi\left(\boldsymbol{r},\boldsymbol{r}_2, ....,\boldsymbol{r}_N\right) \right|^2 \,d\boldsymbol{r}_2.....d\boldsymbol{r}_N
\label{eq3}
\end{eqnarray}
with $\psi\left(\boldsymbol{r},\boldsymbol{r}_2, ....,\boldsymbol{r}_N\right)$, the normalized wave function of the $N$ electron atom. From Eq. (\ref{eq3}) it is clear that $\rho(\boldsymbol{r})$  stands for the normalized single-particle charge density of the many-particle system. In close analogy with (\ref{eq2}), the momentum-space entropy is defined by
\begin{eqnarray}
S_{\gamma}=-\int \gamma(\boldsymbol{p})\, \ln \gamma(\boldsymbol{p})\,d\boldsymbol{p}.
\label{eq4}
\end{eqnarray}
The single-particle momentum-space charge density $\gamma(\boldsymbol{p})$ in Eq.(\ref{eq4}) is characterized by the momentum-space wave function $\phi\left(\boldsymbol{p},\boldsymbol{p}_2, ....,\boldsymbol{p}_N\right)$  obtained by taking the Fourier transform of the position-space wave function. The position- and momentum-space entropies as defined by Eqs. (\ref{eq2}) and (\ref{eq4}) allowed Bialynicki-Birula and Mycielski \cite{ref9} to introduce a stronger version of the uncertainty relation which for a $3$-dimensional system reads
\begin{eqnarray}
S_\rho+S_\gamma\ge 3(1+\ln \pi).
\label{eq5}
\end{eqnarray}
Equation (\ref{eq5}) is known as the BBM inequality. Clearly, this inequality indicates the reciprocity between the representation and momentum spaces such that high values of $S_\rho$  are associated with low values of $S_\gamma$.

For the $N$-electron atom the position- and momentum-space Fisher information are defined by \cite{ref2}
\begin{eqnarray}
F_\rho=\int \frac{1}{\rho(\boldsymbol{r})} \left[\boldsymbol{\nabla} \rho(\boldsymbol{r})\right]^2\,\,d\boldsymbol{r}
\label{eq6}
\end{eqnarray}
and
\begin{eqnarray}
F_\gamma=\int \frac{1}{\gamma(\boldsymbol{p})} \left[\boldsymbol{\nabla} \gamma(\boldsymbol{p})\right]^2\,\,d\boldsymbol{p}.
\label{eq7}
\end{eqnarray}
Understandably, the gradient operators in Eqs. (\ref{eq6}) and (\ref{eq7}) belong to position- and momentum-spaces respectively. Equations  (\ref{eq6}) and (\ref{eq7}) can be written in the equivalent forms \cite{ref10}
\begin{eqnarray}
F_\rho=4\int\left|\boldsymbol{\nabla} \psi(\boldsymbol{r})\right|^2\,\,d\boldsymbol{r}
\label{eq8}
\end{eqnarray}
and
\begin{eqnarray}
F_\gamma=4\int\left|\boldsymbol{\nabla} \phi(\boldsymbol{p})\right|^2\,\,d\boldsymbol{p}.
\label{eq9}
\end{eqnarray}
For computational purposes it will be profitable to work with Eqs. (\ref{eq8}) and (\ref{eq9}) rather than Eqs. (\ref{eq6}) and (\ref{eq7}).

From statistical viewpoint, Fisher information has been realized as a measure of disorder or smoothness of some probability density $P(X)$ and uncertainty of the associated random variable $X$. The disorder aspect has been studied in some length by Frieden \cite{ref11}. The uncertainty properties are clearly delineated by the Stam inequalities \cite{ref12}. The product $F_\rho F_\gamma$  has been conjectured to exhibit a nontrivial lower bound \cite{ref13} such that for $3$-dimensional systems in the $s$ state
\begin{eqnarray}
F_\rho F_\gamma \ge 36.
\label{eq10}
\end{eqnarray}
Equation (\ref{eq10}) is sometimes called a Fisher-based uncertainty relation. As with the BBM inequality which provides a constraint on the allowed values of Shannon position- and momentum-space information entropies, the inequality in Eq.(\ref{eq10}) also sets a limit for the allowed values of Fisher information.

\section{Single-particle charge densities of two-electron systems}
If the effect of correlation in a two-particle system is neglected, then it can be represented by a wave function which is separable in the coordinates of the particles. For example, the well known separable wave function for the Helium atom is given by \cite{ref14}
\begin{eqnarray}
\psi\left(\boldsymbol{r},\boldsymbol{r}_2\right)=c\,e^{-Z(r+r_2)},
\label{eq11}
\end{eqnarray}
where $c$  is a normalization constant and the atomic number $Z$ is a variational parameter. In writing Eq. (\ref{eq11}) we have used atomic units and we shall follow this convention throughout the paper. From  Eqs. (\ref{eq3}) and (\ref{eq11}) we can show that 
\begin{eqnarray}
\rho\left(\boldsymbol{r}\right)=\frac{Z^3}{\pi}\,e^{-2 Z r}
\label{eq12}
\end{eqnarray}
represents the normalized the single-particle position-space charge density of the two-electron atom. The charge density in Eq.(\ref{eq12}) could also be obtained by constructing a normalized single-particle wave function
\begin{subequations}
\begin{eqnarray}
\psi\left(\boldsymbol{r}\right)=\frac{Z^{3/2}}{\sqrt{\pi}}\,e^{- Z r}
\label{eq13a}
\end{eqnarray}
by integrating  Eq. (\ref{eq11}) over the variable $\boldsymbol{r}_2$. One can verify that the suggested alternative approach is equally applicable for the momentum-space ($p$-space) wave function such that
\begin{eqnarray}
\phi\left(\boldsymbol{p}\right)=\frac{2^{3/2}Z^{5/2}}{\pi(p^2+Z^2)^2}.
\label{eq13b}
\end{eqnarray}
\label{eq13}
\end{subequations}
 We shall now follow this approach to find the single-particle wave function and/or charge density for the correlated {two-electron} wave function  that depends explicitly on the inter-electronic separation $r_{12}$. Historically, such wave functions were introduced by Hylleraas \cite{ref15} in 1929. Since then many attempts have been made to write correlated wave functions which can recover a large percentage of electron correlation energy  for both two- and many-electron systems. Recently, Gr\"unels et al \cite{ref16} presented an authentic survey on explicitly correlated electronic structure theory with particular emphasis  on  its application to large systems. On the other hand, Johnson et al \cite{ref17} made use of the  Monte Carlo method to critically examine the performance of different correlation factors as used in the wave functions of small  molecules. In ref. $6$ Mitnik and Miraglia  considered three Hylleraas type wave functions with correlation  factor $F_{12}^{1\,{\rm or}\,2}=1-\mu \,e^{-\lambda\,r_{12}}$ and $F_{12}^{3}=1+\frac{r_{12}}{2} \,e^{-\lambda\,r_{12}}$. We have chosen to work with $F_{12}^{3}$ for studying the role of correlation to modify bare values of some information theoretic quantities. Admittedly, our choice should be based on  some physically founded assumptions. To that end we note that the use of a linear correlation factor $F_{12}^{s}=1+\frac{r_{12}}{2}$ returns more than 80\% of the correlation energy of helium \cite{ref18}, and $F_{12}^{3}$ is only  a simple variant of  $F_{12}^{s}$. Moreover, in addition to satisfying usual correlation cups conditions \cite{ref19a}, $F_{12}^{3}$ has a non-zero second derivative \cite{ref19} and the associated non-separable two-particle wave function becomes separable as $r_{12}\rightarrow \infty$ \cite{ref20}. The linear factor do not satisfy these criteria. For helium, the use of $F_{12}^{3}$ tend to recover correlation energy which is somewhat improved over the corresponding result found by the use of $F_{12}^{1\,{\rm or}\,2}$ \cite{ref6}. Further, it has been pointed out in ref. $17$ that $F_{12}^{3}$ represents one of the highly performing correlation factors. Clearly, these points tend to provide some justification for our choice of the two-particle wave function.

In explicit form the two-particle wave function corresponding to the correlation factor $F_{12}^3$ is given by \cite{ref6}
\begin{eqnarray}
\psi(\boldsymbol{r}_1(=\boldsymbol{r}),\boldsymbol{r}_2,\boldsymbol{r}_{12})=\psi_1(\boldsymbol{r}_1,\boldsymbol{r}_2)+\psi_2(\boldsymbol{r}_1,\boldsymbol{r}_2,\boldsymbol{r}_{12}),
\label{eq14}
\end{eqnarray}
where
\begin{subequations}
\begin{eqnarray}
\psi_1(\boldsymbol{r}_1,\boldsymbol{r}_2)=\frac{NZ^3}{\pi}\cosh(a\,r_1)\cosh(a\,r_2)\,e^{-Z(r_1+r_2)}
\label{eq15a}
\end{eqnarray}
and 
\begin{eqnarray}
\psi_2(\boldsymbol{r}_1,\boldsymbol{r}_2,\boldsymbol{r}_{12})&=&\frac{ N \,\lambda\,Z^3}{2\pi}\cosh(a\,r_1)\cosh(a\,r_2)\nonumber\\&&\,e^{-Z(r_1+r_2)}\,\,r_{12}e^{-b r_{12}}.
\label{eq15b}
\end{eqnarray}
\label{eq15}
\end{subequations}
For $\lambda=0$ and $a=0$ the entangled wave function in Eq.(\ref{eq14}) goes over to the separable one written in Eq.(\ref{eq11}). The explicit $r_{12}$  dependence of $\psi(\cdot)$ appearing in Eq.(\ref{eq15b}) accounts for the  effect of correlation that arises due to inter-electronic repulsion. In constructing the wave function (\ref{eq14}) special care was taken to simulate its behavior at the electron-electron and electron-nucleus coalescence points. The separable product represented by the cos-hyperbolic terms in Eq.(\ref{eq15}) was introduced in order to account for cusp conditions and shielding of electron 2(1) on 1(2). It is well known that cusp conditions lead to a wave function which is continuous but not smooth (discontinuous first derivative) at the singularities or coalescence points.

Equation (\ref{eq15a}) can easily be integrated over $\boldsymbol{r}_{2}$ to get
\begin{eqnarray}
\psi_1(\boldsymbol{r})=\frac{8NZ^4 (Z^2+3 a^2)}{(Z^2-a^2)^3}\,e^{-Z r}\cosh(a\,r).
\label{eq16}
\end{eqnarray}
However, the presence of $r_{12}$  in Eq.(\ref{eq15b}) appears to provide an awkward analytical constraint to carry out a similar integration to get $\psi_2(\boldsymbol{r})$ in simple form. Fortunately, we can make judicious use of the second derivative  of the identity \cite{ref22}

\begin{eqnarray}
\frac{e^{-\mu r_{12}}}{r_{12}}=\frac{1}{2 \pi^2}\int \frac{e^{i\boldsymbol{q}\cdot (\boldsymbol{r}_1-\boldsymbol{r}_2)}}{\mu^2+q^2}\,d\boldsymbol{q}
\label{eq17}
\end{eqnarray}
for separation of variables in Eq.(\ref{eq15b}) and thereby integrate the resulting expression to get
\begin{eqnarray}
\psi_2(\boldsymbol{r})&=&N\lambda Z^3 \frac{e^{-Z r}}{r}\cosh(a\,r) \frac{\partial^3}{\partial \,b^2\,\partial Z}\left(\frac{e^{-(Z+a)r}-e^{-br}}{(Z+a)^2-b^2}\right.
\nonumber\\&+&\left.\frac{e^{-(Z-a)r}-e^{-br}}{(Z-a)^2-b^2}\right).
\label{eq18}
\end{eqnarray}
The complete representation-space single-particle wave function is given by
\begin{eqnarray}
\psi(\boldsymbol{r})=\psi_1(\boldsymbol{r})+\psi_2(\boldsymbol{r}).
\label{eq19}
\end{eqnarray}
Making use of Eqs. (\ref{eq1a}) and (\ref{eq1b}) the Fourier transforms  of Eqs. (\ref{eq16}) and (\ref{eq17}) can be written as  
\begin{eqnarray}
\phi_1(\boldsymbol{p})&=&\frac{4\sqrt{2} N Z^3(Z^2+3a^2)}{\pi^{3/2} (Z^2-a^2)^3}\left(\frac{Z-a}{(Z-a)^2+p^2}\right.\nonumber\\ &+&\left. \frac{Z+a}{(Z+a)^2+p^2}\right)
\label{eq20}
\end{eqnarray} 
and
\begin{eqnarray}
\phi_2(\boldsymbol{p})&=&\lambda N Z^3\sqrt{\frac{2}{\pi}}\,\,\frac{\partial^2}{\partial b^2}\left(\frac{a-Z}{\gamma_m^2}\,X(\delta_m)-\frac{a+Z}{\gamma_p^2} X(\delta_p)\right.\nonumber\\&-&\left.\frac{2}{\gamma_m}\,Y(\delta_m)-\frac{2}{\gamma_p} Y(\delta_p)\right).
\label{eq21}
\end{eqnarray}
Here 
\begin{subequations}
\begin{eqnarray}	
 \gamma_m=(Z-a)^2-b^2 \,\,\,{\rm and}\,\,\, \gamma_p=(Z+a)^2-b^2.
 \label{eq22a}
\end{eqnarray}
The functions $X(\delta_i)$ and $Y(\delta_i)$ are given by
\begin{eqnarray}
X(\delta_i)&=&\frac{1}{p^2+4 Z^2}-\frac{1}{p^2+\alpha^2}-\frac{1}{p^2+\beta^2}\nonumber\\&+&\frac{1}{p^2+\delta_i^2}
\label{eq22b}
\end{eqnarray}
and 
\begin{eqnarray}
Y(\delta_i)=\frac{Z}{(p^2+4Z^2)^2}+\frac{\delta_i}{(p^2+4\delta_i^2)}.
\label{eq22c}
\end{eqnarray}
In  Eqs. (\ref{eq22b}) and (\ref{eq22c}) $i=m$ and $p$ . Also we have
\begin{eqnarray}
\alpha&=&b+a+Z, \,\,\,\beta=b-a+Z,\,\,\nonumber\\
\delta_m&=&Z-a,\,\,{\rm and}\,\,\,\delta_p=Z+a.
\label{22d}
\end{eqnarray}
\label{eq22}
\end{subequations}
Obviously, the complete momentum-space wave function is given by 
\begin{eqnarray}
\phi(\boldsymbol{p})=\phi_1(\boldsymbol{p})+\phi_2(\boldsymbol{p}).
\label{eq23}
\end{eqnarray}
The normalized wave function in Eqs. (\ref{eq19}) and (\ref{eq23}) can now  be used to compute  results for position- and momentum-space single-particle charge densities $\rho(r)$  and $\gamma(p)$ for two-electron atoms. In the wave function of  Mitnik and Miraglia \cite{ref6} we have  $\lambda=1$.  The expressions for $\rho_c(r)$  and  $\gamma_c(p)$ which involve the effect of correlation are obtained my making use of Eqs.(\ref{eq19}) and (\ref{eq23}) respectively. The corresponding charge densities   $\rho_0(r)$  and  $\gamma_0(p)$  of the uncorrelated systems are found in the limit  $\lambda=0$. In Table $1$ we present the results of   $\rho(r=0)$  and  $\gamma(p=0)$ for K-shell electrons of all neutral atoms from helium to neon. For ready reference we display the values of the parameters of the wave function in columns $2$ and $3$ of the table.
\begin{center}
\begin{tabular}{|c|c|c|c|c|c|c|}
\hline
Z&a&b&$\rho _0(r=0)$&$\rho_c( r=0)$& $\!\gamma _0(p\!=\!0)$&$\!\gamma_c(p\!=\!0)$\!\\
\hline
2&0.483802&0.1768&2.1145&1.1952&0.1669&0.1972\\
\hline
3&0.602754&0.3351&7.7567&7.0628&0.0426&0.0475\\
\hline
4&0.712600&0.4916&  18.4685&17.6710&0.0167&0.0181\\
\hline
5&0.776139&0.6630& 36.9494&35.3527	&0.0084	&0.0085\\
\hline
6&0.863445&0.8274&64.5299&62.1643&0.0045&0.0047\\
\hline
7&0.948237&0.9994&103.2280&99.9355&0.0028&0.0029\\
\hline
8&1.032940&1.1592&154.8950&150.4873&0.0018&0.0019\\
\hline 
9&1.116300&	1.3266&	221.4240&215.0098&0.0012&0.0013\\
\hline 
10&1.198190&1.4950&	304.7020&298.1236&0.0008&0.0009\\
\hline 
\end{tabular}
\vskip 0.15cm 
{{\bf Table $1$}: {Single-particle charge densities for K-shell electrons of neutral atoms from helium to neon.}}
\end{center}
\begin{figure}[h]
\centerline{\includegraphics[width=8cm,height=6cm]{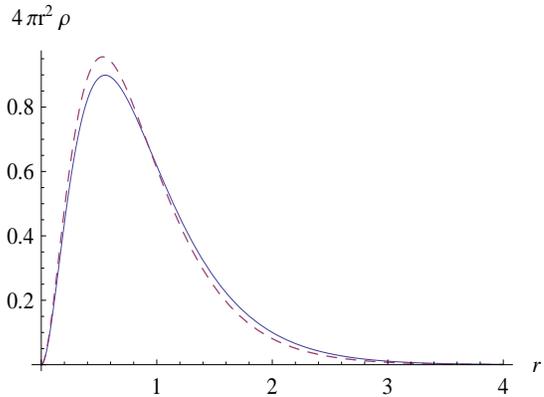}}
\caption{\label{fig1} Single-particle  position-space charge density for the K-shell electrons of helium as a function of $r$. The solid and dashed lines represent variation of densities for the correlated and bare systems respectively.} 
\end{figure}
\begin{figure}[h]
	\centerline{\includegraphics[width=8cm,height=6cm]{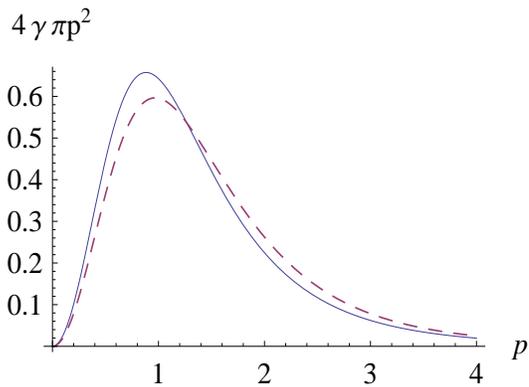}}
	\caption{\label{fig2} Single-particle momentum–space charge density for the K-shell electrons of helium as a function of $p$. The solid and dashed lines carry similar meaning as those in Fig.\ref{fig1}.} 
\end{figure}

The values of $\rho_0(r=0)$ and $\gamma_0(p=0)$  presented here were calculated using the wave functions in Eqs. (\ref{eq16}) and (\ref{eq20}). These wave functions provide some modification over those given in Eq.(\ref{eq13})  because of the factor $\cosh(a\,r_1)\,\cosh(a\,r_2)$  in Eq. (\ref{eq15}). However, for $a=0$  the normalized wave functions obtained from Eqs.(\ref{eq16}) and (\ref{eq20}) coincide with the expression in Eq. (\ref{eq13})  such that $\rho_0^{a=0}(r=0)=0.3183\,Z^3$ and $\gamma_0^{a=0}(p=0)=0.8106/Z^3$. For all values of $Z$  the results for $\rho_0^{a=0}(r=0)$  are greater than those for $\rho_0(r=0)$. This implies that the presence of  the separable cos-hyperbolic term in Eq.(\ref{eq15}) as used to improve on the quality of  the two-particle wave function reduces the value of the single-particle position-space charge density at $r=0$. On the other hand, we observe the opposite for the momentum-space charge density. However, the difference between the numbers for $\rho_0^{a=0}(r=0)(\gamma_0^{a=0}(p=0))$  and  $\rho_0(r=0)(\gamma_0(p=0))$  becomes rather insignificant for large values of the atomic number. Looking closely into the results in Table $1$ we see that for all values of $Z$  the correlated and uncorrelated charge densities satisfy the inequalities $\rho_c(r=0)<\rho_0(r=0)$ and $\gamma_c(p=0) >\gamma_0(p=0)$. Thus we find that the inter-electronic repulsion always causes reduction in the values of position-space electronic charge cloud at $r=0$  and  the opposite is seen to be true for the charge cloud as viewed in the momentum space. However, as expected, in both cases,  the effect of correlation is significant only for low $Z$ values. It will, therefore, be interesting to examine  the overall effect of  inter-electronic repulsion on the coordinate- and momentum-space charge densities for small atomic numbers.

 We display in Fig.$1$ the results for $4\pi r^2 \rho(r)$ as a function of $r$ for $Z=2$. The solid curve denotes the variation of the correlated charge density while the dashed curve represents similar variation of the uncorrelated charge density. From this figure it is clear that inter-electronic  repulsion pushes the bare or uncorrelated charge distribution away from the origin such that values of $\rho_c(r)$ are less than those of $\rho_0(r)$   for $r\le 1$   but for $r>1$  we observe the opposite. We have verified that the deviation between the solid and dashed curves gradually diminishes as we go to higher $Z$ atoms. This is physically realizable since effect of correlation tends to play less dominant role as we move along the periodic table. To visualize the role of correlation in modifying the bare momentum-space charge density we plot in Fig. \ref{fig2} values of  $4\pi p^2 \gamma(p)$ as a function of $p$ . As before, the solid and the dashed curves refer to the variation of correlated and uncorrelated distributions. Rather than being flattened , the bare charge distribution in this case is squeezed due to inter-electronic repulsion such that the values of the correlated charge density are bigger than those of the uncorrelated density for small $p$ values. For large values, however, the solid curve falls below the dashed one.

\section{Results for Shannon entropy and Fisher information}
The single-particle charge densities found from the wave functions (\ref{eq19}) and (\ref{eq23}) can now be used in Eqs. (\ref{eq2}) and (\ref{eq3}), to compute numbers for Shannon information entropies. We present in Table $2$ the results for position- and momentum-space entropies, namely, $S_{\rho}$  and $S_{\gamma}$. The subscripts $0$  and $c$ on $S_{\rho}$  and $S_{\gamma}$  have been used to differentiate between the uncorrelated and correlated values of the entropy.
\begin{center}
\begin{tabular}{|c|c|c|c|c|c|c|}
\hline
Z&$S_{\rho _0}$&$S_{\rho_c}$& $S _{\gamma_0}$& $S_{\gamma_c} $&$S_{\rho _0}+S _{\gamma_0}$&$S_{\rho _c}+S _{\gamma_c}$\\
\hline
2& 2.4323&2.6004&4.1688&3.9158& 6.6011&6.5162\\
\hline
3& 1.0989&1.2391&5.4917&5.2695&6.5906&6.5086\\
\hline
4& 0.1812&0.3099&6.4044&6.1923&6.5856&6.5022\\
\hline
5&-0.5361&-0.4165&7.1172&6.9113&6.5761&6.4948\\
\hline
6&-1.1041&-0.9892&7.6831&7.4498&6.5790&6.4607\\
\hline
7&-1.5812&-1.4702&8.1517&7.9173&6.5705&6.4471\\
\hline
8&-1.9922&-1.8833&8.5686&8.2924&6.5764&6.4091\\
\hline 
9&-2.3534&-2.2467&8.9289&8.63916&6.5755&6.3864\\
\hline 
10&-2.6758&	-2.5748&9.2505&8.9290&6.5747&6.3542\\
\hline 
\end{tabular}
\vskip 0.15cm 
{{\bf Table $2$}: Shannon information entropies of K-shell electrons  of all atoms  in Table $1$}
\end{center}
As in the case of single-particle charge densities presented in Table 1, the position- and momentum-space uncorrelated  entropies, $S_{\rho_0}$  and $S_{\gamma_0}$ carry the effect of improvement sought in the cusp condition of the wave function in Eq.(\ref{eq15}). The uncorrelated information entropies corresponding to $S_{\rho _0}$ and $S _{\gamma_0}$ free from this effect can be calculated by the use of wave functions in Eqs. (\ref{eq13a}) and (\ref{eq13b}) and thus obtain \cite{ref23}
\begin{eqnarray}
S_{\rho}^{a=0}=3-\ln\left(\frac{Z^3}{\pi}\right) \,\,{\rm and}\,\,\nonumber\\S_{\gamma_0}^{a=0}=\ln\left(32 \pi^2 Z^3\right)-\frac{10}{3}.
\label{eq24}
\end{eqnarray}
From Eq.(\ref{eq24}) we see that for the separable wave function (\ref{eq11}), the position-space entropy is a decreasing function of $Z$  such that the values of $S_{\rho_0}^{a=0}$  become negative after $Z=3.9812$. On the other hand, the corresponding momentum-space entropy is an increasing function of $Z$. We note that the $Z$-dependence of Shannon entropies predicted by this simple-minded representation of the two-particle system remains valid even for more realistic description of the system. For example, the numbers for position-space entropies in Table $2$ decrease with $Z$ while those for momentum-space entropies increase with the atomic number independently of whether they are correlated or uncorrelated. But for the change in sign of the position-space entropy, we see that, instead of $Z=3.9812$, both $S_{\rho_0}$ and $S_{\rho_c}$  exhibit zeros at a value of $Z$  between $4$ and $5$. One of our important task here is to examine the role of correlation in modifying the bare entropy values. From our results of  $S_{\rho_0}$ and $S_{\rho_c}$ we see that for all  $Z$  the bare values of the position-space entropy are augmented due to the effect of correlation. As opposed to this, the numbers for  $S_{\gamma_0}$ and $S_{\gamma_c}$  confirm that inter-electronic  correlation reduces the bare values of the momentum-space entropies. Thus correlation plays an opposite role in affecting position- and momentum-space entropies.  As regards the relative importance of correlation in quantifying the position-and momentum-space quantities, one can verify that  bare values of position- space entropies are more susceptible to correlation effect.

From Eq. (\ref{eq24}) we see that the entropy sum $S_{\rho_0}^{a=0}+S_{\gamma_0}^{a=0}$ is $Z$  independent and has a constant value $c_S=6.5665$ . This is, however, not true for the sums  $S_{\rho_0}+S_{\gamma_0}$ and $S_{\rho_c}+S_{\gamma_c}$ shown in columns $6$ and $7$ of Table $2$. The numbers for the entropy sum $S_{\rho_0}+S_{\gamma_0}$  are greater than  $c_S$ for all values of the atomic number and the $Z$ dependence of the sum is such that it exhibits maximum deviation from $c_S$  for $Z=2$  and  tends towards $c_S$  as $Z$ increases. The results for the correlated entropy sum  are less than $c_S$  for all $Z$ and decrease continuously as we go along the periodic table. Some general remarks in respect of this are now in order. For example, the result in Eq.(\ref{eq5}) states that the sum of position- and momentum-space entropies provides a statement for the stronger version of the Heisenberg uncertainty relation.  One would expect this relation  to exhibit some kind of $Z$ dependence. Looking from this point of view the wave function for two-particle systems as given in  Eq.(\ref{eq11}) is physically inadequate.

Making use of the wave functions (\ref{eq19}) and (\ref{eq23}) in Eqs. (\ref{eq8}) and (\ref{eq9})  we computed the numbers for position- and  momentum-space Fisher information. The results for the values of $F_{\rho}$ and $F_{\gamma}$ are presented in Table $3$.  As in the case of Shannon information entropies, the results for $F_{\rho_0}$ and $F_{\gamma_0}$  provide  some modified values over those found from the expression for  $F_{\rho_0}^{a=0}$ and $F_{\gamma_0}^{a=0}$ given by
\begin{eqnarray}
F_{\rho_0}^{a=0}=4Z^2\,\,\,{\rm and}\,\,\,F_{\gamma_0}^{a=0}=12/Z^2.
\label{eq25}
\end{eqnarray}
We have deduced the results in Eq. (\ref{eq25}) by using Eq. (\ref{eq13}) in Eqs. (\ref{eq8}) and (\ref{eq9}). Equation (\ref{eq25}) shows that the position-space Fisher information is an increasing function of $Z$ while the momentum-space one is a decreasing function. The predicted $Z$ dependence of Fisher information by the simple-minded Hartree-type model does not change even for more realistic description of the system. Both uncorrelated and correlated values for position-space Fisher information, $F_{\rho_0}$ and  $F_{\rho_c}$, increase with $Z$. For all values of the atomic number, the results for $F_{\rho_0}$ are greater than those for $F_{\rho_c}$.On the other hand, the corresponding momentum-space quantities decrease with $Z$. In contrast to the position-space information, the numbers for $F_{\gamma_0}$ are less than the results for  $F_{\gamma_c}$. Thus effect of correlation reduces the values of bare position-space information but increases the bare momentum-space results. The momentum-space information is highly sensitive to correlation effect. For $Z=2$ correlation increases the  value of $F_{\gamma_0}$  by about $20\,\%$. As opposed to this, correlation reduces the value of $F_{\rho _0}$  by $8.87\,\%$ only.    

\begin{center}
\begin{tabular}{|c|c|c|c|c|c|c|}
\hline
Z&$F_{\rho _0}$&$F_{\rho_c}$& $F _{\gamma_0}$& $F_{\gamma_c} $&$F_{\rho _0}F_{\gamma_0}$&$F_{\rho _c}F _{\gamma_c}$\\
\hline
2&13.3096&12.1287&3.9434&4.7319&52.8451&57.3918\\
\hline
3& 31.7643& 28.4294&1.6074&1.9505& 51.0579&55.4430\\
\hline
4&58.0412&53.8456& 0.8681&1.0037&50.3856&54.0448\\
\hline
5&  92.8915&86.5378&0.5361&0.8140&49.7991&53.1342\\
\hline
6&135.1810&126.2190&0.3664&0.4183&49.5303&52.7974\\
\hline
7&185.3460&	173.4250&0.2663&0.3025&49.3576&52.4611\\
\hline
8&243.3420&	227.8750&0.2023&0.2299&49.2081&52.3884\\
\hline 
9&309.2030&	264.5170&0.1589&0.1971&49.1324&52.1363\\
\hline 
10&382.9400&359.2810&0.1281&0.1439&49.0546&51.7005\\
\hline 
\end{tabular}
\vskip 0.15cm 
{{\bf Table $3$}: Fisher information of K-shell electrons  of all atoms  in Table $1$.}
\end{center}

From Eq.(\ref{eq25}) we find that the product $F_{\rho_0}^{a=0}\,\, F{\gamma_0}^{a=0}=48$. This result is greater than the lower bound $c_F=36$ of the Fisher-based uncertainty relation (\ref{eq10}). For all values of the atomic number the results for $F_{\rho_c}\,F_{\gamma_c}$ are greater than the corresponding results for $F_{\rho_0}\,\, F{\gamma_0}$. But a common feature of the uncorrelated and correlated products is that both of them start with values greater than $F_{\rho_0}^{a=0}\,\, F{\gamma_0}^{a=0}=48$  and tend towards $48$ as $Z$  increases. Finally, we remark that the $Z$ independence of the product  $F_{\rho_0}^{a=0}\,\, F{\gamma_0}^{a=0}$  results from an oversimplification of the physical reality and, in fact,  the results for $F_{\rho}\,\, F{\gamma}$ should exhibit $Z$ dependence as shown by the numbers in columns $6$ and $7$ of Table $3$.
      
\section{Concluding remarks }
The Hylleraas-type atomic wave functions depend explicitly on the inter-electronic separation. As a result it is rather difficult to use these wave functions in applicative context. In particular, one needs to implement purely numerical routines from the very beginning to apply them in studying physical problems. Given the present-day computer facility this is, however, not a very serious problem. But we believe that there are distinct advantages to viewing problems of physics within the framework of simple analytical models, since many physical effects are then readily expressed and evaluated. Keeping this in view we tried to  construct  analytic expressions for the single-particle charge densities of K-shell electrons of neutral atoms, represented by an  entangled or non-separable wave function and subsequently used them to study the effect of electron-electron correlation on Shannon information entropy ($S$) and Fisher information ($F$). An added realism of our approach is that  we could easily identify the  uncorrelated and correlated parts in the expressions for $S$ and $F$ so as to envisage a detailed study  in respect of the  role of correlation in modifying  the bare values of the Shannon entropy and Fisher information.

The results presented by us in Tables $1$ and $2$  for Shannon entropy and Fisher information clearly show that $Z$ dependence of Fisher information is opposite to that of Shannon entropy. For example, the results for  $F_{\rho} (F_{\gamma})$ increase (decrease) with $Z$  while those for $S_\rho (S_\gamma)$ decrease (increase) with atomic number. These properties of  $S$ and $F$ can be explained simply by using the basic concepts of Shannon entropy and Fisher information. Shannon entropy measures the spatial delocalization of the electron density in an atomic or molecular system while Fishier information provides a measure of localization of the same density distribution. This complementary nature of $S$ and $F$ accounts for their observed variation with respect to the atomic number.

The effect of correlation reduces the values of $F_{\rho_0}$ but increases the results for $S_{\rho_0}$. The opposite is true for the corresponding momentum-space quantities. From our results in Table $2$ we see that the entropic  uncertainty relation  satisfies the inequality  $S_{\rho_0}+S_{\gamma_0}\,>\,S_{\rho_c}+S_{\gamma_c}$. This is true for all values of $Z$. Similarly from the data in Table $3$, we can write $F_{\rho_0}F_{\gamma_0}\,<\,F_{\rho_c}F_{\gamma_c}$  for the Fisher-based uncertainty relation. We would venture to suggest that the observed opposite nature for the response of $S$ and $F$ to inter-electronic repulsion has its origin  in the complementary descriptions of disorder in the density distribution. 



\begin{thebibliography}{1000}
\bibitem{ref1} C. E. Shannon, Bell Syst. Tech. J. \textbf{27}  (1948) 623.
\bibitem{ref2} R. A. Fisher, Proc. Cam. Phil. Soc. \textbf{22} (1925) 7000.	
\bibitem{ref3} V. Vedral, Introduction to Quantum Information Science (Oxford University Press, Oxford(2006)); 
 S. P. Flego , A.  Plastino   and  A. R. Plastino, Int. Res. J. Pure and Appl. Chem. \textbf{2} (2012) 25.
\bibitem{ref4} 4.	P. Ziesche, V. H. Smith Jr., M. Ho, S. P. Rudin, P. Gersdorf and M. Taut, J. Chem. Phys. \textbf{110}  (1999) 6135;
 N. L.  Guevara, R. P. Sagar and R.O. Esquivel, J. Chem. Phys. \textbf{122} (2005) 084101;
 R. P. Sagar and N. I. Guevera, arXiv: Quant-ph//0602046v1 3 Feb 2006 and references therein;
  C. H. Lin and Y. K. Ho, Chem. Phys. Lett. \textbf{633} (2015) 261;
  Aparna Saha, B. Talukdar and Supriya Chatterjee, Physica A \textbf{474} (2017) 370 .
\bibitem{ref5} E. Romera and  J. S. Dehesa, J. Chem. Phys. \textbf{120} (2004) 8906.
\bibitem{ref6} D. M. Mitnik and J. E. Miraglia, J. Phys. B : At. Mol. and Opt. Phys. \textbf{38}  (2005) 3325.
\bibitem{ref7} L. D. Landau and E. M. Lifshitz, Quantum Mechanics- Non-relativistic Theory (Pergamon Press, New York(1975)).
\bibitem{ref8} G. B. Arfken, Mathematical Methods for Physicists, second ed. ( Academic Press, New York (1970)).
\bibitem{ref9} I. Bialynicki-Birula and J. Myceilski, Commun. Math. Phys. \textbf{44} (1975) 129 .
 \bibitem{ref10}  E. Romera, F. S\'anchez-Moreno and J. S. Dehesa, Chem. Phys. Lett. \textbf{414} (2005) 468 .
 \bibitem{ref11} B. R. Frieden,  Science from Fisher information (Cambridge University Press, Cambridge (2004)).
 \bibitem{ref12} A. Stam, Inform. Control. \textbf{2} (1959) 105 .
\bibitem{ref13} J. S. Dehesa, R. Gonzalez-Ferez and Sanchez-Moreno, J. Phys. A : Math. Theor. \textbf{40}  (2007) 1845.
\bibitem{ref14} L. I. Schiff, Quantum Mechanics Tata McGraw Hill (New Delhi, India(2010)).
\bibitem{ref15} E. A. Hylleraas, Z. Phys. \textbf{54}  (1929) 347
\bibitem{ref16} A. Gr\"unels, S Hirata, Y. Ohnishi and S. Ten-no, J. Chem. Phys. \textbf{146} (2017) 080901. 
\bibitem{ref17} C. M. Johnson,  S. Hirata and S. Te-no, Chem. Phys. Lett. \textbf{683} (2017)247. 
\bibitem{ref18} W. Kutzelnigg, Theor. Chim. Acta {\bf 68} (1985)445.
\bibitem{ref19a}T. Kato, Commun. Pure Appl. Math. \textbf{10} (1957)151; C. C. J. Roothaan and A. W. Weiss, Rev. Mod. Phys. \textbf{32} (1960) 194.
\bibitem{ref19} W. Klopper, F. R. Manby, S. Te-no and E. F.Veleev, Int. Rev. Phys. Chem. {\bf 25} (2006)427.
\bibitem{ref20} H. J. Monkhorst, Mol. Phys. {\bf 103} (2005) 2009.
\bibitem{ref22} S. Bhattacharyya, A. Bhattacharya, B. Talukdar and N. C. Deb, J. Phys. B : At. Mol. Opt. Phys. \textbf{29} (1996) L147.
\bibitem{ref23} N. L. Guevara, R. P. Sagar and R. O. Esquivel, Phys. Rev. A \textbf{67} (2003) 012507.
\end{thebibliography}
\end{document}